\documentclass[aps,prb,twocolumn,groupedaddress,showpacs]{revtex4}
\usepackage[dvips]{graphics}

\begin{document}

\title{Critical Susceptibility Exponent Measured from Fe/W(110)
Bilayers}

\author{M.J. Dunlavy} 
\email{dunlavmj@mcmaster.ca}
\author{D. Venus}

\affiliation{Dept. of Physics and Astronomy, McMaster University, 1280
Main St. West, Hamilton Ontario, Canada}

\date{\today}

\begin{abstract}
The critical phase transition in ferromagnetic ultrathin Fe/W(110)
films has been studied using the magnetic ac susceptibility.  A
statistically objective, unconstrained fitting of the susceptibility
is used to extract values for the critical exponent $\gamma$, the
critical temperature T$_{c}$, the critical amplitude $\chi_{o}$ and
the range of temperature that exhibits power-law behaviour.  A fitting
algorithm was used to simultaneously minimize the statistical variance
of a power law fit to individual experimental measurements of
$\chi$(T).  This avoids systematic errors and generates objective
fitting results.  An ensemble of 25 measurements on many different
films are analyzed.  Those which permit an extended fitting range in
reduced temperature lower than approximately $4.75\times10^{-3}$
give an average value $\gamma$=1.76 $\pm$0.01.  Bilayer films give a
weighted average value of $\gamma = 1.75\pm0.02$.  These results are
in agreement with the 2-dimensional Ising exponent $\gamma$=
$\frac{7}{4}$.  Measurements that do not exhibit power-law scaling as
close to T$_{c}$ (especially films of thickness 1.75ML) show a value
of $\gamma$ higher than the Ising value.  Several possibilities are
considered to account for this behaviour.

\end{abstract}


\maketitle

\section{Introduction}
\label{intro}

Experiments that seek to measure critical phase transitions are very
important to physics at a fundamental level.  Careful experiments can
be used to confirm or deny the theoretical models of universality and
scaling.  The true nature of the order parameter of a system, both in
terms of dimensionality and degrees of freedom, is revealed at the
transition and important physical insight is gained in the looking.

An ultrathin magnetic film closely approaches the physical realization
of a truly two dimensional magnetic system, and offers a better system
for studying critical phase transitions in two dimensions than more
traditional layered bulk materials such as
Rb$_{2}$CoF$_{4}$ \cite{rb2cof4}, where interlayer interactions will
always be present, even if only to a small degree.  Bander and Mills
 \cite{prb_bander} have shown that when ferromagnetic thin films have
uniaxial anisotropy, the critical regime near the Curie temperature is
described by the 2 dimensional Ising model.  For this reason, a great
number of measurements of the static critical exponents of ultrathin
ferromagnetic films have been reported.  Almost all of this
experimental work concentrates on the critical exponent of the
magnetization, $\beta$ \cite{prb_elmers, prl_li92, nature_back,
apa_ballentine89, prb_huang94, prb_li90, jap_elmers96,
prb_kohlhepp92}.  To our knowledge, there are only a handful of
reports in which the critical exponent of the magnetic susceptibility,
$\gamma$, is investigated experimentally for an ultrathin magnetic
film \cite{jmmm_bovensiepen, prb_elmers, jap_elmers96, rsi_arnold97,
prb_rudt, prb_stetter}.  Unfortunately, almost all of these
susceptibility studies have at least one of a number of deficiencies
which call the results into question.

A common difficulty in the determination of critical exponents is the
determination of T$_c$.  Small variations in the assumed value of
T$_c$ have a profound effect on the fitted value of the critical
exponent, and introduce confidence limits that are usually much larger
than those derived from a simple two parameter fit for the critical
exponent and amplitude.  The extreme sensitivity of the results to
T$_c$ implies that the same data must be used to determine both T$_c$
and the critical parameters.  This is particularly true for metastable
ultrathin films, since very small shifts in the critical temperature
are often introduced by temperature cycling and annealing, or by
residual vacuum contamination.  A second difficulty is determining the
temperature range where scaling is observed.  Since real, finite
systems do not show infinite divergences, the order parameter departs
from power law behaviour close to T$_c$ because of finite size
effects, dynamical effects (in ac measurements), a finite
demagnetization factor, and so on.  To fit the data properly, an
objective, four parameter power-law fit of the data is required.  In
addition to T$_{c}$ and $\gamma$, the fit should find values for the
critical amplitude $\chi_{o}$ and the cutoff for power-law behaviour
near T$_{c}$.  Finally, in order to demonstrate true systematic
behaviour, it is clear that the analysis of many films and
measurements is necessary.

It is perhaps surprising that after two decades of investigating the
critical properties of ultrathin ferromagnetic films, that no
published measurements of $\gamma$ meet these criteria.  An impressive
study by Back et al. \cite{nature_back} on Fe/W(110) ultrathin films
determines $\beta$ and the exponent of the critical isotherm $\delta$
using the dc magnetization, and then derives $\gamma$ using the
scaling relations between different exponents.  The value of T$_c$ is
not fit, but rather taken to lie at the peak of the dc susceptibility
for a particular experiment. The results agree with the predictions of
the 2D Ising model.  This represents a check of the internal
consistency of the data and scaling relations, but is not an
independent measurement of $\gamma$.  Elmers et al. \cite{prb_elmers}
report dc susceptibility results for a series of submonolayer films of
Fe/W(110) and find $\gamma$ = 2.8 $\pm$ 0.2, significantly different
than the 2D Ising value of 1.75 \cite{pr_onsager}.  It is not clear to
what extent this finding is a result of using an incomplete film layer
or if, as they suggest, the material is exhibiting the behaviour of an
anisotropic Heisenberg system.  Other studies report results only for
a single measurement from a single film \cite{rsi_arnold97}.  Still
others use questionable criteria for determining T$_c$, such as the
disappearance of the imaginary component of the susceptibility in an
ac measurement  \cite{prb_rudt}, the peak of the real ac
susceptibility \cite{prb_stetter, prb_garreau}, or the presence of a
``shoulder" above the peak of the
susceptibility \cite{jmmm_bovensiepen} under special circumstances.

This paper presents the results of a collection of 25 measurements of
the ac magnetic susceptibility of Fe films between 1.5 and 2.0ML grown
epitaxially on W(110), and the values of $\gamma$ derived from them
using an objective minimization of the statistical variance between
the data and a power law fit using four parameters: T$_c$, $\gamma$,
the amplitude $\chi_0$, and the low reduced-temperature cutoff t$_x$
for fitting.  Error estimates on T$_c$ and $\gamma$ are provided by
1$\sigma$ variations in the statistical $\chi{^2}$.  The results fall
into two distinct classes.  Measurements exhibiting power law
behaviour over a long range of reduced temperature extending down to a
cutoff $t_{x} < 4.75\times10^{-3}$ give an average critical exponent
$\gamma$ = 1.76 $\pm$ 0.01.  Measurements which exhibit power law
behaviour down to larger values of t$_x$ show a systematic trend to
higher values of $\gamma$ which depends rather linearly on ln(t$_x$).
The possibility that films which give a high value of $\gamma$ have a
distribution of transition temperatures will be addressed to explain
this unexpected result.

\section{Theory}
\label{theory}
According to scaling theory, the real component of the intrinsic
magnetic susceptibility ($\chi_{int} = \partial{M}/ \partial{H}$) above
the Curie temperature of a critical phase transition is described by
the power-law equation:
\begin{equation}
 \chi_{int}^{'}(t) = \chi_{o} t^{-\gamma}
\label{gamma}
\end {equation}
where $\chi_{o}$ is the critical amplitude, $\gamma$ is the static
critical exponent for the susceptibility of the order parameter and
$t$ is the reduced temperature above T$_{c}$, given as:
\begin{equation}
t = (\frac{T - T_{c}}{T_{c}}).
\label{red_T}
\end{equation}

For real measurements of the magnetic ac susceptibility, additional
terms need to be added to account for both demagnetization and
dynamical effects.  The demagnetizing factor N is folded into the
expression for the intrinsic susceptibility by augmenting the magnetic
field by:
\begin{equation}
H_{eff} = H - NM
\label{demag_expression}
\end{equation}
where $H_{eff}$ is the effective field acting on the ferromagnet.
This gives rise to an effective susceptibility of:
\begin{equation}
\chi^{'}_{eff}(T) = \frac{\chi^{'}_{int}(T)}{1+N\chi^{'}_{int}(T)}.
\label{demag_suscept}
\end{equation}
It is easy to see that for a non-zero value of $N$, the susceptibility
cannot diverge at T$_{c}$.  $N$ will ``dampen'' any experimental
measurement of $\chi$ as long as the value of the product
$N\chi_{int}$ is comparable to or greater than one.

To accurately describe results from ac susceptibility, it is necessary
to add the effect of the relaxation time of the magnetization to the
effective susceptibility.  In the linear response approximation, for
systems with an exponential relaxation time ($\tau$) [M($\mathcal{T}$)
$\propto$ exp(-$\mathcal{T}$/$\tau$)] where $\mathcal{T}$ is time under
the influence of an externally applied sinusoidal field, the real
dynamic susceptibility ($\chi'$) can be written as:
\begin{equation}
\chi^{'}(T) = \frac{\chi^{'}_{eff}(T)}{1 + (\omega\tau(T))^{2}}
\label{dynamic_susceptibility}
\end{equation}
where $\omega$ is the driving frequency of the magnetic field
\footnote{In these equations, the relaxation time of the magnetization
is used, where $\frac{dM}{dt} = -\frac{1}{\tau}(M-M_{\infty})$.
Dissipation can also be expressed in terms of the relaxation of the
effective field using the Landau-Liftshitz equation and the damping
parameter.  The two terms are related by $\tau =
\frac{\chi_{eff}}{\lambda}$.}.  This final form of the magnetic
susceptibility limits the ability of experiments to probe critical
behaviour very close to the transition.  To observe any critical
scaling in the experimental data, two requirements must be met: we
must have $N$ sufficiently small and we must have $(\omega\tau)^{2}
<< 1.0$.

The extreme aspect ratio of ultrathin films leads to very small values
of $N$.  For systems that have their moments oriented in-plane, $N$ is
proportional to first order to the thickness divided by the effective
lateral dimension of the film \cite{chikazumi}.  For films that are one
or two atomic layers thick and many thousands of lattice spacings
wide, $N$ will be extremely small.  This is another reason why
ultrathin films are ideal for studies of critical phenomenon in two
dimensions.  Previous studies of the susceptibility on ultrathin films
have attempted to estimate $N$ (and include the estimation in the
power-law fits) by using the maximum value of the real susceptibility
 \cite{prb_rudt}.  The argument proceeds by rearranging
eq.(\ref{demag_suscept}) as follows:
\begin{equation}
\frac{1}{\chi_{eff}(T)} = \frac{1}{\chi_{int}(T)} +N
\label{bad_suscept}
\end{equation}
This leads one to the conclusion that at T$_{c}$, when $\chi_{int}$ is
infinite, $N$ = 1/$\chi_{max}$.  This simple treatment has several
problems even for dc susceptibility measurements (where $\omega$ = 0)
in that it ignores other effects (finite field, saturated correlation
length, etc.) that will saturate the susceptibility and will give a
value for $N$ that is artificially too high and is at best an upper
limit \cite{jmmm_aspelmeier}.  If this limit of $N$ was then used in
the power-law analysis, the resulting quoted values for $\gamma$
should be called into question.

Dynamic effects are only significant near T$_{c}$ where critical
slowing down will lead to a large relaxation time for the
equilibration of the order parameter near  \cite{rmp_hohenberg}.  This
can be less of a problem in dc measurements, but the increased
signal-to-noise that is achieved in ac measurements make the effort to
deal with the dynamics problem one worth undertaking. In fact,
critical slowing effects should disappear once the temperature is
increased more than a degree or two above T$_{c}$.  Dynamic effects
will change the temperature at which the susceptibility exhibits a
maximum (depending on the measurement frequency used), making the
evaluation of T$_{c}$ by that method difficult if not impossible.

\section{Experiment}
\label{experiment}

Fe/W(110) ultrathin films with high quality epitaxial layers can be
grown at least up to 2 ML \cite{ss_waller}.  Previous studies of
Fe/W(110) have shown that the magnetic properties of the films depend
sensitively on the film thickness \cite{prl_elmers95}.  There have been
studies made on films less than 1.5ML that show interesting
perpendicular magnetic behaviour due to the film structure which
results from step-flow growth \cite{apl_hauschild, prb_elmers99}.  For
this study, we will concentrate on the thickness range from 1.5 to 2.0
ML so as to ensure in-plane magnetic behaviour.

The experiments were performed in an UHV environment with a base
pressure of 1x10$^{-10}$torr.  The films were grown by molecular beam
epitaxy (MBE) from a 99.995\% pure iron wire.  The substrate was a
tungsten single crystal that had been cut and polished to expose the
[110] face.  The cut is accurate to within 0.4$^{\circ}$.  The first layer
was deposited at room temperature and then annealed for one minute to
500K.  This slight annealing produces increased sharpness of the
resulting pseudomorphic LEED pattern.  Further depositions were
performed at room temperature with no annealing.  The film growth,
thickness and quality were monitored by Auger electron spectroscopy
(AES) and LEED.

\begin{figure}
\begin{center}
\scalebox{.5}{\includegraphics{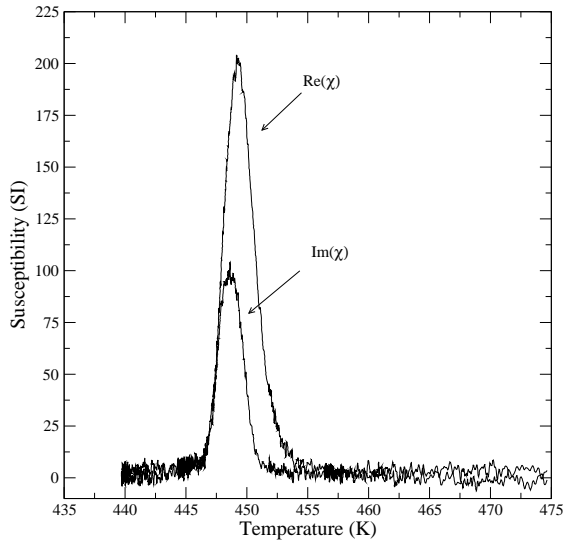}}
  \caption{Magnetic ac-susceptibility measured from a 1.8ML film of
    iron grown upon W(110).  The real and imaginary components of the
    susceptibility were measured simultaneously.}
\end{center}
\label{typical_data}
\end{figure}

The ac-$\chi_{m}$ measurements were made via the surface magneto-optic
Kerr effect (SMOKE) using a focussed He-Ne laser spot with a diameter
of approximately 0.75mm.  Small coils near the surface produced a
sinusoidally oscillating magnetic field, H, which influences the
moments in the paramagnetic film above T$_{c}$.  The field was applied
along the film's easy axis [1$\overline{1}$0].  The surface
magneto-optic Kerr effect produces a rotation of the polarization of
the laser light reflected off of the magnetic surface.  The signal
manifests itself in changes in the light intensity at the photo-diode
detector placed in front of the reflected laser beam.  The 1f signal
is read by a dual-phase lock-in amplifier that can simultaneously
record both the in-phase (or real) susceptibility ($\chi^{'}$(T)) and
the out-of-phase (or imaginary) susceptibility ($\chi^{''}$(T)).  The
raw signal is calibrated to SI units and the entire signal can be
represented as
\begin{equation}
\chi(T) = \chi^{'}(T) + i \chi^{''}(T).
\label{complex_suscept}
\end{equation}
Fig.(\ref{typical_data}) shows a typical measurement of the complex
susceptibility measured from a 1.8ML iron film.  The measurement was
made with an applied field amplitude of 0.7 Oersteds at a frequency of
400Hz.

\begin{figure}
\scalebox{.5}{\includegraphics{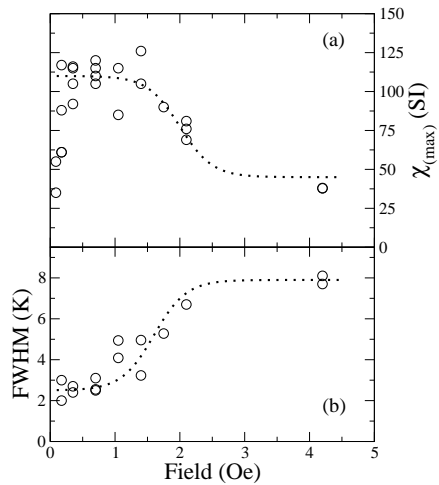}}
\caption{(a) The maximum value of the magnetic susceptibility as a
function of applied magnetic field.  (b) FWHM of the real
susceptibility peak plotted as a function of applied field amplitude.
The minimum half-width is achieved for fields less than 1 Oersted.}
\label{fwhm}
\end{figure}

It would be best to use an infinitesimally small field, but of course
this is not possible experimentally. A study of magnetic
susceptibility peak shape as a function of field was conducted to see
what value of the field would give the best compromise between signal
and finite field effects.  It showed that the `peak' shape of the
susceptibility data measured in field amplitudes of 1.0 Oe and lower
is not sensitive to the size of the field.  Fig.(\ref{fwhm}a) and
(\ref{fwhm}b) show the maximum value and full-width, half-maximum
(FWHM) for the susceptibility peaks respectively as a function of the
amplitude of the applied field.  The trend below 1.0 Oe in both graphs
is stable, (except at extremely low fields where the signal itself
disappears) but deviates for higher fields.  Resulting measurements of
a susceptibility peak measured in these small fields give a FWHM
typically between two and three and a half degrees.In these
measurements, smaller field amplitudes were accessible but this
generally lead to a degradation of the signal-to-noise ratio.

Sample heating was accomplished by running ac current (no more than 1A
rms) through a small tungsten wire filament located behind the
tungsten crystal.  AC-current at 60Hz was used to reduce the effects
of stray offset fields at the surface.  It had been found in the past
that a dc-current introduced a 0.1 Oe offset field at the surface.
The 0.1Oe field caused by the heating filament is much less than the
applied field used in the measurement (typically 0.7Oe) and is much
less than the field which increases the FWHM of the susceptibility
peak (Fig.\ref{fwhm}). Any questions about the effect of the heating
current were answered by comparing data taken while increasing and
decreasing the film temperature respectively.  The value of current
used in the two methods differed by a factor of three, and there was
absolutely no difference in the final data.  The temperature of the
film was measured using a W/WRh thermocouple embedded in the tungsten
crystal and the rate of temperature increase/decrease was in most
cases limited to 0.2deg/min.  This low rate more than adequately
compensates for thermal variations in the crystal and permits even
heating of the film over the entire surface (app. 1cm$^{2}$).

\section{Data Analysis}
\label{data_analysis}
To fit the susceptibility data to eq.(\ref{gamma}), an objective,
many-parameter fit was used to determine the best values for the Curie
temperature, T$_{c}$, the critical exponent $\gamma$, the critical
amplitude, $\chi_{o}$, and $t_{x}$ which is the smallest value of
the reduced temperature to show power law scaling. 

The fit is performed in double logarithm space [ln($\chi$) vs ln(t)],
the slope of which will correspond to the critical exponent.  Taking
the logarithm of the susceptibility necessitated the removal of data
points where $\chi$(T) goes to zero.  Since these points are weighted
the least in the fits [weighting in logarithmic space goes as
(1/$\chi_{d}^{2}$)], it is felt that this ``weeding'' out of points
does not adversely affect the final fit.  A small range of
temperatures close to the peak was chosen for possible best values of
T$_{c}$ used in the reduced temperature.  For each considered value of
$T_{c}$, a weighted least-squares fit was performed on the data in the
new log-log data space from ln($t_{max}$) (which always corresponds to
the data point measured at the highest temperature) to a cutoff value
ln($t_{x}$).  

\begin{figure}
\scalebox{.5}{\includegraphics{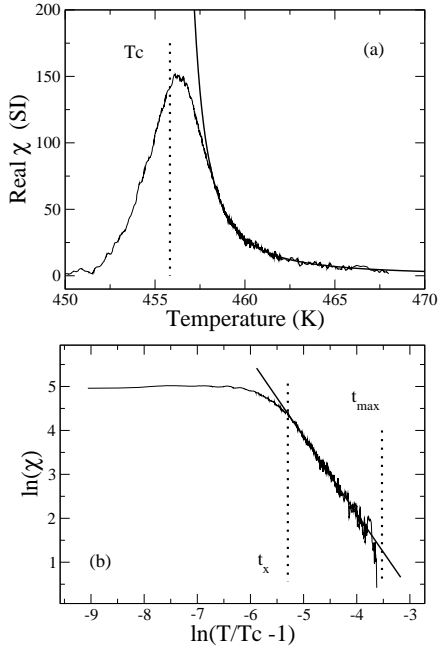}}
\caption{Power law fit for a typical susceptibility measurement. 
(a) $\chi$ versus temperature. Solid line shows the fit and dotted line
shows position of T$_{c}$.  (b) Fit in log-log space, with dotted
lines showing position of t$_{max}$ and t$_{x}$. t$_{max}$ always
corresponds to the maximum temperature which was measured.}
\label{typical_fit}
\end{figure}

\begin{figure}
\scalebox{.5}{\includegraphics{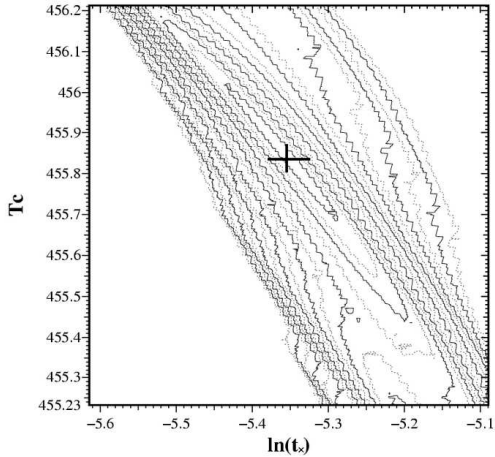}}
\caption{Contour plot of s$^{2}$ as a function of T$_{c}$ and
ln($t_{x}$).  The global minimum (indicated by the cross) shows the
values of T$_{c}$=455.84 and ln($t_{x}$)=-5.355 corresponding to the
best fit.}
\label{fitting_results}
\end{figure}

$t_{x}$ was itself varied over a range from just below $t_{max}$ to a
value of $t$ where the power-law scaling was obviously no longer
valid.  The variance of the fit was minimized for the best value of
$T_{c}$ and the cut-off, $t_{x}$.  The variance is the best test for a
fit made in a many-parameter space \cite{bevington} where the number of
points does not remain constant.  It is given by:
\begin{equation}
s^{2} =
\sum_{i_{t_{max}}}^{i_{t_{x}}}\frac{(ln(\chi_{i})-F(t_{i}))^{2}}
{\sigma_{i}^{2}}/\sum_{i_{t_{max}}}^{i_{t_{x}}}{\frac{1}{\sigma_{i}^{2}}}
\label{variance}
\end{equation}
where $\chi_{i}$ is the $i^{th}$ data point, $F(t_{i})$=ln$(\chi_{o})
+\gamma ln(t)$ is the fitted function, and $\sigma_{i}$ is the error
associated with the logarithm of each data point.  The fitting
algorithm generates a list of fitting results for all possible
combinations of the inputs for the data in
fig.(\ref{typical_fit}). Fig.(\ref{fitting_results}) shows a contour
plot of s$^{2}$ as a function of T$_{c}$ and ln(t$_{x}$).  There is a
global minimum at T$_{c}$ = 455.84K and ln($t_{x}$) = -5.355
(corresponding to a temperature of 457.99K). There are local minima
exhibited in the graphs that have higher values for $t_{x}$ than the
global minimum.  The fact that the global minimum fits the data closer
to T$_{c}$ increases its significance.

\begin{figure}
\scalebox{.5}{\includegraphics{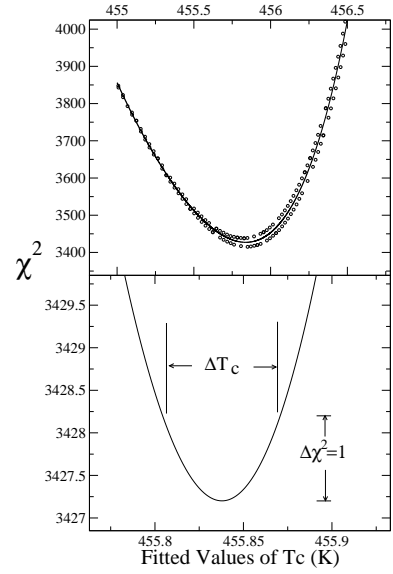}}
\caption{Graphs of $\chi^{2}$ versus value of T$_{c}$ used in fit.
Fig.(a) shows the minimum in $\chi^{2}$ with a smooth function fit to
the points.  Fig.(b) shows the fitted curve with an indicated range
corresponding to a change in $\chi^{2}$ of 1.0.  The value of T$_{c}$
with error for this data set is T$_{c}$ = 455.84$\pm$.03K}
\label{err_tc}
\end{figure}

To get an error estimation on T$_{c}$, the fits were recalculated
while keeping the optimum value of ln($t_{x}$)=-5.36 to allow for a
careful statistical $\chi^{2}$ analysis for a consistent number of
data points.  According to statistics for a multi-variable
fit \cite{bevington}, the 65\% confidence range for a fit value is
judged by whatever range of the parameter corresponds to the limit
where the value of the unreduced $\chi^{2}$ increases by one.
Fig.(\ref{err_tc}) shows $\chi^{2}$ versus T$_{c}$ for the data in
fig.(\ref{typical_fit}).  Due to the good signal-to-noise of the data
and the large number of points in the limited temperature range, the
error for T$_{c}$ is very small.  The number of points in the fit used
for Fig.\ref{err_tc} is 1905, which will give a reduced $\chi^{2}$
for the fit of 1.8, signifying a very good fit to the data.  The
T$_{c}$ value from this analysis is 455.84$\pm$0.03.  While this
range of T$_{c}$ creates an uncertainty in $\gamma$ on the order of
the error from the least-square analysis, the two effects should
compound to increase the confidence limit on $\gamma$ slightly.  The
value for the critical exponent from the particular data set in
fig.(\ref{typical_fit}a) is $\gamma=1.75\pm0.02$.  The fitted critical
amplitude $\chi_{o}$ is $7.3\pm0.3\times10^{-3}$.

It now becomes necessary to check for both dynamic and demagnetization
effects in the data.  It has already been remarked that a
demagnetization factor equal to 1/$\chi_{max}$ provides an upper limit
on the value of $N$.  This assumption would lead to a value of $N$ for
the data in fig.(\ref{typical_fit}a) to be about 1/150 or
6.67x10$^{-3}$.  According to eq.(\ref{demag_suscept}), once the value
of the product $N\chi$ approaches 0.05, the observed power law
behaviour of the intrinsic susceptibility is lost.  For this data set,
this would occur at a temperature of 464.2K.  This would give a value
of ln(t$_{x}$) approximately equal to -4.0.  In other words, if we
believe the above estimate for $N$, then no linear segment in
double-log space would extend closer to T$_{c}$ than this.  The
results in fig.(\ref{typical_fit}b) clearly show the linear segment
extending much lower than -4.0.  The value of $N$ must therefore be
much smaller.  The power-law behaviour in fact deviates at a
temperature of approximately 458.0K.  If we take the `5\% rule' a step
further, the maximum value of $N$ then becomes approximately 1/1632 or
6.1x10$^{-4}$, a full order of magnitude lower than the previous
estimate.  This lower value is more in keeping with the value of $N$
expected from geometric arguments and provides a new upper limit on
$N$.

Checking the saturation from dynamic effects requires a more definite
knowledge of the time response of the moments as a function of
temperature than these measurements currently allow.  However, a
simple calculation can be made on a theoretical basis.  Near T$_{c}$,
the relaxation time of the magnetization will undergo critical slowing
down which, by theory, follows the formula:
\begin{equation}
\tau(T) = \tau_o t^{-z\nu}
\label{tau}
\end{equation}
where $\nu$ is the critical exponent associated with the correlation
length and $z$ is the critical slowing down exponent.  While there are
very few experiments that measure the critical slowing down of the
relaxation time, $\tau$, on ferromagnetic systems, theoretical
simulations \cite{prb_lacasse93, prb_wang93} suggests that the value of
$z$ should be approximately 2.2 for the 2D Ising system.  The value
for $\tau_o$ should be very small, on the order of inverse-GHz to
agree with FMR resonance frequencies.

To see no dynamic effect interference in $\chi_{int}$ as per
eqn.(\ref{dynamic_susceptibility}), you need $\omega\tau$ less than
1.0.  If we apply the `5\%' rule again and use $z\nu$ = 2.2, $\tau_o$
= 1x10$^{-9}$s and $\omega$=(2$\pi$)150.0Hz, we find that ln($t_{x}$)
will be -5.6.  This is close to the fit value for ln($t_{x}$) and may
be the reason for the saturation of the susceptibility.  Better
estimates of $\tau_{o}$ and $z\nu$ are required to pursue this
question further.

\section{Results from Many Films}
\label{many_films}
\begin{figure}
\scalebox{.5}{\includegraphics{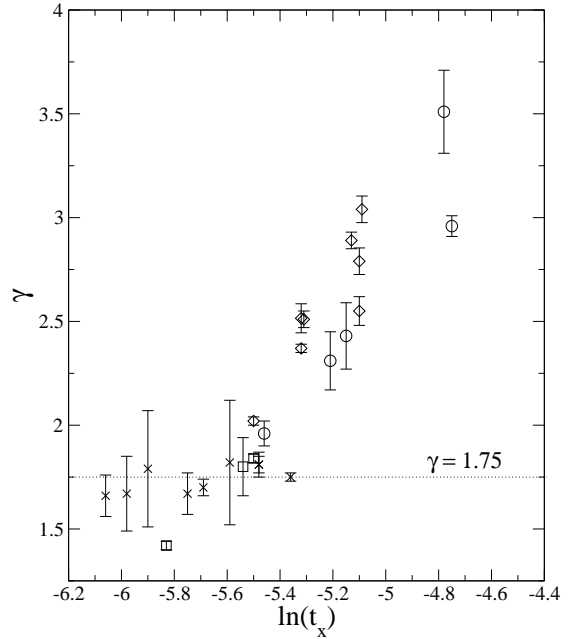}}
\caption{Best fit values of $\gamma$ plotted as a function of reduced
temperature cutoff, ln(t$_{x}$).  Open circles represent films that
are slightly less than 1.5ML, squares are 1.5ML films, diamonds are
1.75ML, and X's are 2.0ML.}
\label{cutoff_gamma}
\end{figure}

Critical power-law fitting was performed on a sample of 25 different
measurements from many films grown between 1.5 and 2.0 ML.  The
results show that the value of $\gamma$ that comes from the fit is
dependent on the thickness of the film.  Fig.(\ref {cutoff_gamma})
shows a plot of $\gamma$ as a function of ln($t_{x}$) for all 25
measurements.  It is apparent from the plot that $\gamma$ is not only
thickness dependent but that there is also the presence of an extra
effect which causes films with a higher value of $\gamma$ to fit
$t_{x}$ further from T$_{c}$.

The following weighted and unweighted average results for $\gamma$ can
be given: (1) For bilayer films, the weighted average value of
$\gamma$ is 1.75$\pm$0.015 with an unweighted average of 1.74$\pm$0.023
(2) For sesqilayer films, the weighted average is 1.63$\pm$0.01.
(This weighted average is suspect as there are only 3 data points with
small individual error which do not overlap).  The unweighted average
is 1.68$\pm$0.13. (3) The weighted average value of gamma for films
with ln(t$_{x}$) less than -5.35 is 1.76$\pm$.01.  The unweighted
average is 1.76$\pm$0.04.  Most of the films with values of t$_{x}$ in
this last range are either 2.0 or 1.5ML, but it should be stated that
there is also one measurements at 1.75ML and one other just below
1.5ML.

Films with a thickness of 2.0ML and 1.5ML consistently have the lowest
values of ln(t$_{x}$) and these are the films that give (on average)
the 2D Ising result.  There is no supporting evidence to the idea that
all of these measurements would show Ising behaviour if the linear fit
was performed on a range of data closer to T$_{c}$. An examination of
the fitted value of $\gamma$ as a function of t$_x$ for individual
measurements show that there is absolutely no increase in $\gamma$
with increasing t$_x$. It is thought that the higher values of t$_{x}$
are an indicator of another as-yet not understood process that affects
the power law scaling when the film thickness is just below or equal
to 1.5 and 2.0ML.  It is possible that this process is also
responsible for the high value of $\gamma$ reported for films of 0.8ML
thickness \cite{prb_elmers}.

We have examined several possible explanations for this behaviour.
The first involved using corrections to scaling
arguments \cite{goldenfeld, prb_lumsden98} that should be taken into
account for fitting data far away from T$_{c}$. If this is the case,
then the effective value of the exponent, $\gamma_{eff}$ is
approximated by
\begin{equation}
\gamma_{eff} = \gamma - a \Delta |\overline{t}|^{\Delta},
\label{correction}
\end{equation}
where $a$ is a constant, $\overline{t}$ is some ``average
temperature'' representing the fitting range, and the exponent
$\Delta$ is close to 0.5 for Ising systems \cite{goldenfeld} regardless
of the dimensionality of the system.  If $t_x$ is chosen as
$\overline{t}$, the data in fig.(\ref{cutoff_gamma}) can be reasonably
described by eq.(\ref{correction}) with $\Delta \approx 3$.  The large
discrepancy between the fitted and theoretical value of $\Delta$
suggests that corrections to scaling are not the important factor
here.

Another possibility for the rising value of $\gamma$ is the idea of
dimensional crossover from an Ising system to an anisotropic
Heisenberg system as a function of temperature \cite{prb_binder76}.  It
is well known that true two-dimensional Heisenberg systems cannot
support long range magnetic order above T=0 \cite{prl_mermin66}, but
only a small amount of anisotropy is required to lift this
restriction \cite{prb_bander}.  It is possible that the films that
exhibit high values of ln(t$_{x}$) are showing higher values of
$\gamma$ due to a change in the anisotropy. This explanation was also
offered by Elmers et.al. \cite{prb_elmers} for their 0.8ML results, and
it is interesting to note that the value of ln(t$_{x}$) in their
result would be -5.3, which is consistent with the onset of high
$\gamma$ values in this study.  However, there are several reasons
against this idea.  First, none of the data with an Ising exponent
show a crossover to larger $\gamma$ when $t_{x}$ is artificially
increased.  Second, none of the data used in this work, including
those data sets that fit with a ln(t$_{x}$) value less than -6, shows
anything resembling a 'break point' in the double-log slope indicating
different critical power law over different temperature ranges, which
would provide a clear indication that dimensional crossover is
occurring.  Finally, a reduced anisotropy should result in a change in
the trend of the transition temperature as a function of
thickness \cite{prb_erickson91}, an effect which we do not observe.

The third possibility is that the films in the sensitive thickness
range have a wide distribution of transition temperatures.  This
effect on the fitted slope is easy to understand.  If some fractional
area of the film undergoes a phase transition at a temperature
slightly above the average ``mean'' value of T$_{c}$ used to reduce
the temperature for the logarithmic plot, then those areas will
register as an artificially high slope in the fit.  While the exact
nature of the distribution is unknown, it is certain that any
distribution with values above the T$_{c}$ used in the fitting routine
will increase the fitted exponent.  To gauge the effect
quantitatively, a series of data sets were modeled using a normalized
Gaussian distribution of T$_{c}$ and an intrinsic value of $\gamma$ of
1.75.  It was found that it takes a half-width of just over 0.5K to
cause a 1\% increase in $\gamma$.  No significant increase is found to
occur as long as the half-width is less than 0.25K.  A half-width of
1K gives a fit exponent of 1.81, a 3.5\% increase.  To achieve a fit
value for $\gamma$ of 3 (near the maximum fit value in the 25
measurements we fit), requires a half-width of 2.5K.  It may be that
the films less than 1.5ML and between 1.5 and 2.0ML are more sensitive
to small structural inhomogeneities that give rise to a wider
distribution of T$_{c}$.  Films with a complete second monolayer will
be more homogeneous than films that are slightly thinner.  We
speculate that the distribution of transition temperatures may be
related to the distribution of atoms that are located at step edges
between the first and the incomplete second monolayer.  For the
complete 2ML, the films should be very homogeneous and a narrow
distribution may be expected.  The 1.5ML films have equal areas that
are 1ML and 2ML thick respectively and as such present a uniform
configuration of steps which have been shown \cite{prb_elmers99} to
give a correlated magnetic state.  The 1.75 films are on the threshold
of the percolation limit of the second monolayer and it is possible
that slight structural deviations are more likely to cause a wider
distribution of transition temperatures.  This suggestion may also go
towards explaining the 0.8ML results quoted above \cite{prb_elmers}.

\section{Conclusion}
\label{conclusion}

We report the results of fitting measurements of the magnetic
ac-susceptibility for critical power law exponents.  We find that the
critical exponent for bilayer Fe/W(110) films to b
e 1.75$\pm$0.02 and
for films in general with a value of t$_{x}$ below 4.75x10$^{-3}$,
$\gamma = 1.76 \pm 0.01$.  This result confidently place this system
in the 2D Ising universality class.  The fitting routine allows the
simultaneous extraction of the critical exponent and the critical
temperature from a single measurement of the susceptibility.  There is
evidence of another process which affects fitting of the
susceptibility for certain thicknesses.  This may be due to these
films having a larger distribution of critical temperatures.

\begin{acknowledgments}
The authors wish to acknowledge the many technical contributions made
by Marek Kiela.  This work was supported by the National Science and
Engineering Research Council of Canada.  

\end{acknowledgments}

\end{document}